\documentclass[12pt]{article}
\usepackage{graphicx}
\usepackage{color}
\usepackage{cite}

\def \beq{\begin{equation}}
\def \eeq{\end{equation}}
\def\eqref#1{(\ref{#1})}
\def\bea{\begin{eqnarray}}
\def\eea{\end{eqnarray}}

\def\URLtilde{\lower0.2em\hbox{$\tilde{\phantom{a}}$}}
\def\mycomm#1{\hfill\break\strut\kern-3em{\color{red}\tt ====> #1
\color{black}}\hfill\break}

\newcount\timecount
\newcount\hours \newcount\minutes  \newcount\temp \newcount\pmhours
\hours = \time
\divide\hours by 60
\temp = \hours
\multiply\temp by 60
\minutes = \time
\advance\minutes by -\temp
\def\hour{\the\hours}
\def\minute{\ifnum\minutes<10 0\the\minutes
\else\the\minutes\fi}
\def\clock{
\ifnum\hours=0 12:\minute\ AM
\else\ifnum\hours<12 \hour:\minute\ AM
\else\ifnum\hours=12 12:\minute\ PM
\else\ifnum\hours>12
\pmhours=\hours
\advance\pmhours by -12
\the\pmhours:\minute\ PM
\fi
\fi
\fi
\fi
}

\def\monthname{\relax\ifcase\month 0/\or January\or February\or
March\or April\or May\or June\or July\or August\or September\or
October\or November\or December\else\number\month/\fi}
\def\today{\monthname~\number\day, \number\year}
\def\bold#1{\setbox0=\hbox{$#1$}     \kern-.025em\copy0\kern-\wd0
\kern.05em\copy0\kern-\wd0
\kern-.025em\raise.0433em\box0 }
\def\draft{
\color{red}
\noindent
$\bold{\hbox{\tt \Huge Draft:  \clock, \today.}}$
\hfill\break
\vrule width 0pt height 4.5ex
\kern6em
$\bold{\hbox{\tt \Huge Not for distribution.}}$
\par\noindent\color{black}
\vskip-9ex\strut
}

\begin{document}
\setcounter{footnote}{1}
\rightline{EFI 18-11}
\rightline{TAUP 3033/18}
\vskip1.5cm

\centerline{\large \bf Scaling of P-wave excitation energies in
heavy-quark systems}
\bigskip

\centerline{Marek Karliner$^a$\footnote{{\tt marek@proton.tau.ac.il}}
 and Jonathan L. Rosner$^b$\footnote{{\tt rosner@hep.uchicago.edu}}}
\medskip

\centerline{$^a$ {\it School of Physics and Astronomy}}
\centerline{\it Raymond and Beverly Sackler Faculty of Exact Sciences}
\centerline{\it Tel Aviv University, Tel Aviv 69978, Israel}
\medskip

\centerline{$^b$ {\it Enrico Fermi Institute and Department of Physics}}
\centerline{\it University of Chicago, 5620 S. Ellis Avenue, Chicago, IL
60637, USA}
\bigskip
\strut

\begin{center}
ABSTRACT
\end{center}
\begin{quote}
A simple regularity in anticipating P-wave excitation energies 
     of states with heavy quarks is noted.
It can apply to systems such as the negative-parity $\Sigma_c$, $\Sigma_b$,
and $\Omega_c$, $\bar Q Q$ quarkonia, and the bottom-charmed meson $B_c$.
When one subtracts a term
accounting for phenomenological energies of heavy quarks binding with one
another in S-waves, the residual excitation energies display an approximately
linear behavior
in the reduced mass of constituents, all the way from the $\Lambda$ to the
$\Upsilon$.
\end{quote}

\smallskip

\leftline{PACS codes: 12.39.Hg,12.39.Jh,14.20.Lq,14.20.Mr}
\bigskip

\section{Introduction \label{sec:intro}}

The LHCb experiment, working at the CERN Large Hadron Collider, has observed a
number of new baryons containing heavy quarks, including a series of five
excited $\Omega_c = css$ resonances \cite{Aaij:2017nav} and a new $\Xi_b^-
= bsd$ resonance \cite{Aaij:2018yqz}. These have been interpreted, though not
uniquely, as, respectively, P-wave excitations of the ground state $\Omega_c$
\cite{Karliner:2017kfm,Padmanath:2017,Wang:2017vnc,Wang:2017zjw,Chen:2017gnu,%
Aliev:2017led} and one or more P-wave excitations of the ground state
$\Xi_b^-$ \cite{Chen:2018orb}.  We seek simple methods for confirming these
assignments.  Furthermore, it has been of interest to estimate the P-wave
excitation energies for $\Sigma_c$ and $\Sigma_b$ states \cite{Karliner:2015ema}
as well as for the $B_c = b \bar c$ system (see, e.g., \cite{Eichten:1994gt,%
Ebert:2002pp,Kiselev:1994rc,Fulcher:1998ka,Mathur:2018epb}).

Spurred by these developments, we asked whether there is a simple way of
estimating P-wave excitation energies without the use of the two-body or
three-body Schr\"odinger equation, its relativistic analogue, or other
methods such as lattice quantum chromodynamics.  To our surprise, there
appears to be an approximate method which, while not perfect, probably
suffices as a guideline to whether a given state is a P-wave candidate.

The method builds upon a constituent-quark treatment which was used to predict
successfully \cite{Karliner:2014gca} the mass of the $\Xi_{cc}^{++} = ccu$
baryon subsequently discovered by LHCb \cite{Aaij:2017ueg}.  Account was taken
of quark masses, hyperfine interactions, and S-wave binding terms $B(q_1q_2)$
involving any quark pairs where one quark is heavier than $u,d$ and the 
other heavier than $s$.  These binding terms are obtained phenomenologically
by comparing masses of hadrons containing a single heavy quark (e.g., $q_1$
or $q_2$) with ones containing two heavy quarks (e.g., $q_1 \bar q_2$). We find
that when these binding terms are taken into account in calculating S-P mass
differences, the residual energy differences $\Delta E_R$ depend approximately
linearly on the reduced mass $\mu_{12} = m_1 m_2/(m_1 + m_2)$ of the pair.
This behavior extends from the $\Lambda = uds$ baryon all the way up to the
$\Upsilon(1S)$ and their respective P-wave excitations.

We lay out the tools for our estimates in Sec.\ \ref{sec:tools}, describing
assumed quark masses and binding terms.  The ground rules for quoting S-P
splittings are also given.  We quote the observed S-P splittings for a number
of pairs in Sec.\ \ref{sec:SPspl}.  The effects of binding terms, if any, are
considered in Sec.\ \ref{sec:DE}, giving rise to residual energy differences
$\Delta E_R$ which are plotted as functions of reduced mass.  An approximately
linear dependence is seen.  In baryonic cases the problem is reduced to a
two-body one by assuming one quark is excited with respect to two others which
remain in a relative S-wave.

The linear dependence of $\Delta E$ on reduced mass is used in Sec.\
\ref{sec:pred} to predict several quantities which were only crudely estimated
before.  These include P-wave excitation energies for $\Sigma_c$ and
$\Sigma_b$ states \cite{Karliner:2015ema} and for $\Omega_c$ states
\cite{Karliner:2017kfm}.  Predictions for $\Xi_b$ and $B_c$ are also
given and compared with others in the literature.  Section \ref{sec:disc}
is devoted to a discussion of the possible source of the observed regularity,
and a brief conclusion.

\section{Tools \label{sec:tools}}

We use separate constituent-quark masses for mesons and baryons
\cite{Karliner:2014gca}.  They are summarized in Table \ref{tab:qm}.  Analysis
of S-wave mesons and baryons makes use of binding terms $B(q_1 q_2)$, also from
Ref.\ \cite{Karliner:2014gca}, summarized in Table \ref{tab:bin}.  These
terms were calculated by comparing the masses of spin-averaged S-wave bound
states (e.g., for charmonium) with the sum of their constituent-quark masses
as determined from hadrons containing a single heavy quark (e.g.,~$\Lambda_c$).

\begin{table}
\caption{Quark masses in MeV used in this analysis.
\label{tab:qm}}
\begin{center}
\begin{tabular}{c c c} \hline \hline
Quark & In a meson & In a baryon \\ \hline
\vrule width 0pt height 2.5ex 
$u,d$ & $m^m_{u,d} = 310$ & $m^b_{u,d} = 363$ \\ 
 $s$  &    $m^m_s = 483$  &   $m^b_s = 538$ \\
 $c$  & $m^m_c = 1663.3$  & $m^b_c = 1710.5$ \\
 $b$  & $m^m_b = 5003.8$  & $m^b_b = 5043.5$ \\ \hline \hline
\end{tabular}
\end{center}
\end{table}

\begin{table}
\caption{Pair binding terms $B(q_1q_2)$ in MeV used in this analysis.
\label{tab:bin}}
\begin{center}
\begin{tabular}{c c c} \hline \hline
Pair $q_1 q_2$ & $B(q_1q_2)$ & $B(q_1 \bar q_2)$ \\ \hline
\vrule width 0pt height 2.5ex 
$cs$ &  $35.0$ & $70.0$ \\
$bs$ &  $41.8$ & $83.6$ \\
$cc$ &  $129$  & $258$ \\
$bc$ & $170.8$ & $341.5$ \\
$bb$ & $281.4$ & $562.8$ \\ \hline \hline
\end{tabular}
\end{center}
\end{table}

\section{S-P splittings \label{sec:SPspl}}

\subsection{Baryons}

Unless otherwise specified, we take all masses from the 2018 Particle Data
Group listings \cite{PDG18}.  We consider baryons with excitation of a
spinless (scalar) diquark except in the case of $\Omega_c = css$, where we
consider the spin-1 $ss$ diquark to be excited by one unit of orbital
angular momentum with respect to the charmed quark~\cite{Karliner:2017kfm}.

We take the masses listed in Table \ref{tab:bar} to calculate the
spin-averaged S-P splittings shown.  The masses of excited states are
calculated using averages $\bar M_P$ weighted by $2J+1$ factors, where $J$ is
the spin of the resonance.  Small uncertainties in masses are not quoted.

\begin{table}
\caption{Masses of ground state baryons and their orbital excitations
$\Delta E_{P-S}$, in MeV.  Here $\Delta E_{P-S}$ denotes the difference
between spin-weighted average P-wave and S-wave masses.
\label{tab:bar}}
\begin{center}
\begin{tabular}{c c c c c c} \hline \hline
State   & $1/2^+$ & $1/2^-$ & $3/2^-$ & $\bar M_P$ & $\Delta E_{P-S}$ \\ \hline
\vrule width 0pt height 2.5ex 
 $\Lambda$ & 1115.683 & 1405.1  & 1519.5  & 1481.37 & 365.68 \\
$\Lambda_c$ & 2286.46 & 2592.25 & 2628.11 & 2616.16 & 329.70 \\
$\Lambda_b$ & 5619.60 & 5912.20 & 5919.92 & 5917.35 & 297.75 \\
$\Xi_c$ & 2469.37$^a$ & 2792.2$^a$ & 2818.4$^a$ & 2809.6 & 340.3 \\
$\Omega_c^{~b}$ & 2742.33 & \multicolumn{2}{c}{See note $^c$} &
3079.94 & 337.61 \\ \hline \hline
\end{tabular}
\end{center}
\leftline{$^a$Error-weighted isospin average.}
\leftline{$^b$ Spin-averages of ground state and assumed P-wave states from
Ref.\ \cite{Karliner:2017kfm}.}
\leftline{$^c$(2,2,1) states with $J=(1/2,3/2,5/2)$,
cf. Ref.\ \cite{Karliner:2017kfm}.}
\end{table}

\subsection{Mesons}

We consider only those systems for which the spin-averaged ground state and
P-wave masses can be calculated.  They are $c \bar s$ (``$D_s$''), $c \bar c$,
and $b \bar b$.  For $c (\bar u, \bar d)$ not all candidates for the 1P level
are firmly established, while for $b (\bar u, \bar d)$ a spin-zero meson and
one of two predicted spin-1 mesons are still missing (see Sec.\ V B).  For
$b \bar c$ (``$B_c$'') no P-wave states have been seen, but
their masses have been predicted (see Sec.\ V F).  The relevant
masses are shown in Table \ref{tab:mes}.  Spin averaged masses are
\beq
\bar M_S \equiv \left[M(^1S_0) + 3M(^3S_1)\right]/4~,~~
\bar M_P \equiv \left[M(^3P_0) + 3M(^3P_1) + 3M(^1P_1) + 5M(^3P_2)\right]/12~.
\eeq

\begin{table}
\caption{Masses of ground state mesons and their orbital excitations,
in MeV.
\label{tab:mes}}
\begin{center}
\begin{tabular}{c c c c c c c c c c} \hline \hline
State & $M(^1S_0)$ & $M(^3S_1)$ & $\bar M_S^{~a}$ & $M(^3P_0)$ & $M(^3P_1)$ &
 $M(^1P_1)$ & $M(^3P_2)$ & $\bar M_P^{~b}$ & $\Delta E_{P-S}$ \\ \hline
\vrule width 0pt height 2.5ex 
$D_s$ & 1968.34 & 2112.2 & 2076.2 & 2317.7 & 2459.5$^c$ & 2535.1$^c$ & 2569.1 &
 2512.3 & 436.0 \\
$c \bar c$ & 2983.4 & 3096.9 & 3068.5 & 3414.71 & 3510.67 & 3525.38 & 3556.17 &
 3525.3 & 456.8 \\
$b \bar b$ & 9399.0 & 9460.3 & 9445.0 & 9859.44 & 9892.78 & 9899.73 & 9912.21 &
9899.7 & 454.8 \\ \hline \hline 
\end{tabular}
\end{center}
\leftline{$^a$Spin-averaged ground state mass.  $^b$Spin-averaged P-wave mass.}
\leftline{$^c$Orthogonal mixtures of $^3P_1$ and $^1P_1$ states.}
\end{table}

\section{Residual energy differences $\Delta E_R$ \label{sec:DE}}

We now calculate residual energy differences $\Delta E_R \equiv \Delta E_{P-S}
-\sum B$ for the above systems,
where $\sum B$ denotes the sum of $B(q_1q_2)$ over all relevant heavy quarks
$q_1$ and $q_2$ (cf. Table \ref{tab:bin}).  The results are shown in Table
\ref{tab:der}.  Here $[q_1q_2]$ denotes a spinless color-antitriplet diquark,
while $(ss)$ denotes a spin-1 color-antitriplet diquark.  We quote
isospin-averaged masses where appropriate, letting $q$ stand for $u$ or $d$.

\begin{table}
\caption{Residual energy differences $\Delta E_R$ and corresponding reduced
masses, in MeV
\label{tab:der}}
\begin{center}
\begin{tabular}{c c c c c c c c c} \hline \hline
\vrule width 0pt height 2.5ex 
System & $q_1$ & $q_2$ & $m_1$ & $m_2$ & $\mu_{12}$
 & $\Delta E_{P-S}$ & $\sum B$ & $\Delta E_R$ \\ \hline
\vrule width 0pt height 2.5ex 
$\Lambda$ & $[ud]$ & $s$ & 576.0 & 538 & 278.2 & 365.7 & 0 & 365.7 \\
$\Lambda_c$ & $[ud]$ & $c$ & 576.0 & 1710.5 & 430.9 & 329.7 & 0 & 329.7 \\
$\Lambda_b$ & $[ud]$ & $b$ & 576.0 & 5043.5 & 517.0 & 297.8 & 0 & 297.8 \\
 $\Xi_c$ & $[qs]$ & $c$ & 799.8 & 1710.5 & 545.0 & 340.3 & 35.0 & 305.3 \\
 $\Omega_c$ & $(ss)$ & $c$ & 1098.8 & 1710.5 & 669.0 & 337.6 & 70.0 & 267.6 \\
$D_s$ & $c$ & $s$ & 1663.3 & 483 & 374.3 & 436.0 & 70.0 & 366.0 \\
 $c \bar c$ & $c$ & $c$ & 1663.3 & 1663.3 & 831.6 & 456.8 & 258.0 & 198.8 \\
 $b \bar b$ & $b$ & $b$ & 5003.8 & 5003.8 & 2501.9 & 454.8 & 563 & --108.2 \\
\hline \hline
\end{tabular}
\end{center}
\end{table}

\begin{figure}
\begin{center}
\includegraphics[width=0.96\textwidth]{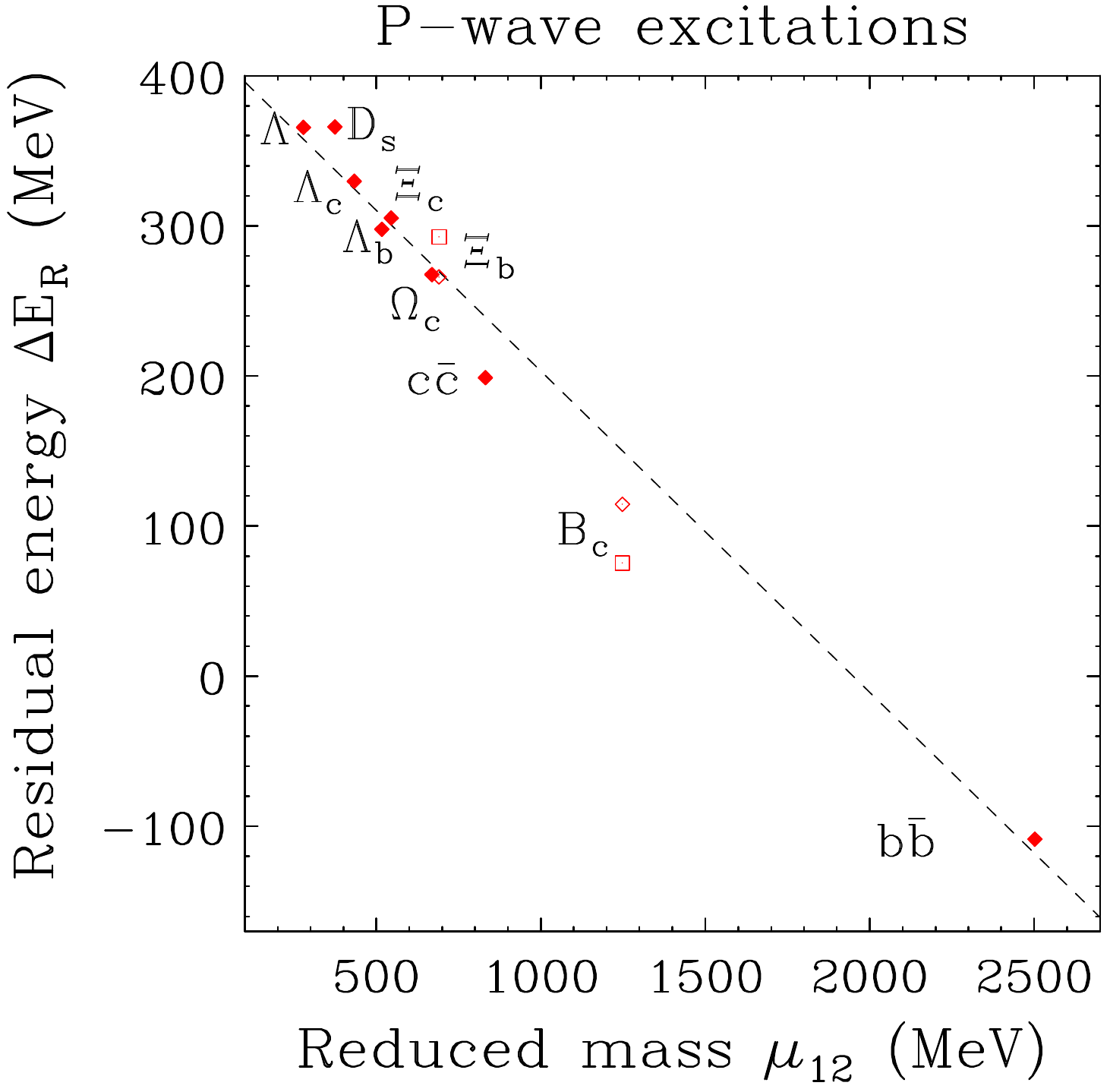}
\end{center}
\caption{Residual energies $\Delta E_R$ as functions of reduced mass $\mu_{12}$.
Dashed line:  fit of Eq.\ (\ref{eqn:lin}).  Filled diamonds denote data used in
the fit.  Theoretical predictions for $\Xi_b$ (subsection E) and $B_c$
(subsection F) systems are plotted as hollow diamonds and hollow squares.
\label{fig:lin}}
\end{figure}

Whereas the quantities $\Delta E_{P-S}$ are not monotonic functions of the
reduced mass $\mu_{12}$, when the binding energies $B$ are subtracted from
them, the residual energies $\Delta E_R$ are crudely arranged along a
straight line, as shown in Fig.\ \ref{fig:lin}.  A linear fit to the 
eight experimentally known values in Table \ref{tab:der} 
gives the result
\beq \label{eqn:lin}
\Delta E_R = (417.37 - 0.2141\, \mu_{12})~{\rm MeV}~.
\eeq
The root-mean-square deviation of the data from this fit is 18.7 MeV.
We discuss some consequences of this regularity, if it is to be taken
seriously, in the next Section.

\section{Consequences and Predictions \label{sec:pred}}

\subsection{$\Sigma_c$ and $\Sigma_b$ baryons}

In Ref.\ \cite{Karliner:2015ema} a linear extrapolation of excitation
energy was used to estimate the S-P wave splittings for $\Sigma_c$ and
$\Sigma_b$ baryons.  The present discussion gives support to that
assumption.  The parameters of the present linear fit give slightly
different values of $\Delta E_R$, as shown in Table \ref{tab:sig}.
For the states in this table, there are no $B$ terms, so $\Delta E_{P-S} =
\Delta E_R$.

\begin{table}[h]
\caption{Values of $\Delta E_R=\Delta E_{P-S}$ predicted by linear fit 
of Eq.\ (\ref{eqn:lin})
compared with those of Ref.\ \cite{Karliner:2015ema}.
\label{tab:sig}}
\begin{center}
\begin{tabular}{c c c c c c c c} \hline \hline
   State & $q_1$ & $q_2$ & $m_1$ & $m_2$ & $\mu_{12}$ & 
 \multicolumn{2}{c}{$\Delta E_R$} \\
         &       &       & (MeV) & (MeV) & (MeV) &
Ref.\ \cite{Karliner:2015ema} & Eq.\ (\ref{eqn:lin}) \\ \hline
\vrule width 0pt height 2.5ex 
$\Sigma$   & $(uu)$ & $s$ & 776 & 538    & 317.7 & 357.5 & 349.3 \\
$\Sigma_c$ & $(uu)$ & $c$ & 776 & 1710.5 & 533.8 & 290.9 & 303.1 \\
$\Sigma_b$ & $(uu)$ & $b$ & 776 & 5043.5 & 672.5 & 238.8 & 273.4 \\
\hline \hline
\end{tabular}
\end{center}
\end{table}

\subsection{Charm and bottom mesons}

The reduced masses for $D$ and $B$ mesons are displayed in Table \ref{tab:dbp}.
They lead to predictions via Eq.\ (\ref{eqn:lin}) of $\Delta E_R$, which is
equal to $\Delta E_{P-S}$ because the binding correction $B$ is zero.
\begin{table}[h]
\caption{Calculation of $\Delta E_R$ and $\Delta E_{P-S} = \Delta E_R +
B(q_1q_2)$ for $D$, $B$, and $B_s$ mesons,
based on linear fit of Eq.\ (\ref{eqn:lin}).  Masses in MeV.
\label{tab:dbp}}
\begin{center}
\begin{tabular}{c c c c c c c c c} \hline \hline
State & $q_1$ & $q_2$ & $m_1$  & $m_2$ & $\mu_{12}$ & $\Delta E_R$
 & $B(q_1q_2)$ & $\Delta E_{P-S}$ \\ \hline
 $D$  & $c$ & $q$  & 1663.3 & 310 & 261.3 & 361.4 & 0 & 361.4 \\
 $B$  & $b$ & $q$  & 5003.8 & 310 & 291.9 & 354.9 & 0 & 354.9 \\
$B_s$ & $b$ & $s$  & 5003.8 & 483 & 440.5 & 323.1 & 83.6 & 406.7 \\
\hline \hline
\end{tabular}
\end{center}
\end{table}

In order to compare these predictions with experiment, one must know the masses
of all four P-wave states.  Our partial information is summarized in Table
\ref{tab:dbo}.  The $D$ and $B$ mass eigenstates have $j$ (the vector sum of
light-quark spin and orbital angular momentum) equal to 1/2 or 3/2.  Those with
$j=3/2$ (total $J=1,2$) decay predominantly via D waves, are narrow, and are
firmly established \cite{PDG18}.  Those with $j=1/2$ ($J=0,1$) are expected to
decay via S waves and are very broad, with consequent mass uncertainty.
The $j=1/2$ $D$ mesons would satisfy the linear fit if their widths, exceeding
200 MeV, were included as error bars.  No candidates for the $j=1/2$ $B$ states
have been identified.  They would have to be considerably lighter than the
$j=3/2$ states if they were to obey the prediction in Table \ref{tab:dbp}.
The outlier nature of $D$ and $B$ states is further discussed 
in Sec.~\ref{sec:disc}.

\begin{table}[h]
\caption{Masses for calculating S-P splitting in charmed and bottom
mesons.  Error-weighted averages over charge states unless otherwise indicated.
\label{tab:dbo}}
\begin{center}
\begin{tabular}{c c c c c c c c c}  \hline \hline
State & $M(^1S_0)$ & $M(^3S_1)$ & $\bar M_S^{~a}$ & $\bar M_P^{~b}$& $M(^3P_0)$
 & \multicolumn{2}{c}{$M(J=1)$} &$M(^3P_2)$ \\
    &    &   &        & (pred.) &       &  $j=1/2$  &  $j=3/2$  & \\ \hline
$D$ &1867.24 & 2008.56 & 1973.23$^a$ & 2334.6 & 2349.2$^c$ & Note $^d$
 & 2420.9 & 2461.1 \\
$B$ & 5279.48 & 5324.65 & 5313.36$^a$ & 5668.2 & ?? & ?? & 5726.0
 & 5738.4 \\
$B_s$ & 5366.89 & 5415.4 & 5403.3$^a$ & 5810.0 & ?? & ?? & 5828.63
 & 5839.85 \\ \hline \hline
\end{tabular}
\end{center}
\leftline{$^a$Spin-averaged ground state mass.  
$^b$Spin-averaged P-wave mass predicted from Eq.~\eqref{eqn:lin}.}
\leftline{$^c$Error-weighted isospin average width 235.7 MeV}
\leftline{$^d$Neutral candidate: $M=2427 \pm40$ MeV, $\Gamma=384^{+130}_{-110}$
 MeV.}
\end{table}

The predicted spin-averaged P-wave mass for $B_s$ is low enough that the
$j=1/2$ $B_s$ P-wave states are probably below the respective $BK$ and $B^*K$
thresholds for the $J=0$ and $J=1$ states.  Thus, like the $D_{s0}(2317)$ and
$D_{s1}(2460)$ (see below), they are expected to be very narrow, decaying only
via $B_{s0} \to \gamma B_s^*$ and $B_{s1} \to \gamma B_s~{\rm or}~\gamma B_s^*$,
or with isospin-violating processes involving $\pi^0$ emission.
The properties of these states have been discussed in Refs.\
\cite{Bardeen:2003kt,Nowak:2003ra}.

\subsection{$D_s$ mesons}

The observed masses of $D_{s0}(2317)$ and $D_{s1}(2460)$ were considerably
below predictions of potential models, leading to some initial surprise.  The
present regularity (Fig.\ \ref{fig:lin}) indeed supports the picture of these
states as lying below those predictions.
\subsection{$\Omega_c$ baryons}

The residual energy $\Delta E_R$ for the five narrow $\Omega_c$ states observed
by LHCb \cite{Aaij:2017nav} lies right on the linear fit, supporting their
assignment as five P-wave states \cite{Karliner:2017kfm,Padmanath:2017,%
Wang:2017vnc,Wang:2017zjw,Chen:2017gnu,Aliev:2017led} and disfavoring an
alternate assignment (see, e.g., \cite{Karliner:2017kfm}) in which the two
highest states are 2S excitations and two lower-mass P-wave states remain to be
discovered.

\subsection{$\Xi_b$ baryons}

\begin{table}[h]
\caption{Values of $\Delta E_{P-S}$ and $\Delta E_R$ for $\Xi_b$ states from 
models compared with predictions of linear fit (\ref{eqn:lin}).  Masses in MeV.
\label{tab:xib}}
\begin{center}
\begin{tabular}{c c c c c c} \hline \hline
$1/2^+$ & $1/2^-$ & $3/2^-$ & $\bar M_P$ & $\Delta E_{P-S}$ & $\Delta E_R$ \\
\hline
\vrule width 0pt height 2.5ex 
5792.19$^a$ & 6096$^b$ & 6102$^b$ &  6100  & 307.8 & 266.0 \\
            & 6120$^c$ & 6230$^c$ & 6126.7 & 334.5 & 292.7 \\
 & \multicolumn{3}{c}{Calculated from Eq.\ (\ref{eqn:lin})} & 311.4 & 269.6 \\
\hline \hline
\end{tabular}
\end{center}
\leftline{$^a$Spin-averaged ground state mass.  $^b$Ref.\ \cite{Chen:2018orb}.
$^c$Ref.\ \cite{Ebert:2011kk}.}
\end{table}
In Table \ref{tab:xib} we compare a recent prediction \cite{Chen:2018orb} for
the masses of P-wave excitations of the scalar $[sq]$ quark in $\Xi_b$ baryons
(Fig.\ \ref{fig:lin}, hollow diamond), with an earlier one (\cite{Ebert:2011kk},
Fig.\ \ref{fig:lin}, hollow square), and with the result of the linear fit
for reduced mass $\mu_{bc}=690.3$ MeV.
The fit is more consistent with the later prediction.

\subsection{$B_c$ states}

One can obtain a value of$\Delta E_{P-S}$ for the $B_c$ system by interpolating
between the nearly equal values for the $c \bar c$ and $b \bar b$ systems, as
one might expect if the interquark potential is close to the logarithmic one
proposed in Ref.\ \cite{Quigg:1977dd}.  One thus obtains $\Delta E_{P-S} = 456$
MeV, corresponding to the open diamond in Fig.\ \ref{fig:lin} when a binding
term of 341.5 MeV is taken into account.  An early potential-model prediction
\cite{Eichten:1994gt} was $\Delta E_{P-S} = 417$ MeV, corresponding to the open
square in Fig.\ \ref{fig:lin}. Subsequent calculations of $\Delta E_{P-S}$ gave
430, 427, and 427 MeV in Refs.\ 
\cite{Ebert:2002pp,Kiselev:1994rc,Fulcher:1998ka}, respectively. The prediction
of Eq.\ (\ref{eqn:lin}), using a reduced mass of $m_cm_b/(m_c+m_b) = 1248.3$
MeV, is $\Delta E_R = 150.1$ MeV, or $\Delta E_{P-S} = 491.6$ MeV, considerably
larger than any of the above values. 

\section{Discussion and conclusions \label{sec:disc}}

The binding terms $B$ used to calculate $\Delta E_R$ represent corrections to
the picture of spectra due to constituent-quark masses and hyperfine terms
\cite{DeRujula:1975qlm}, when quarks are heavy enough to experience the
short-distance Coulomb-like force of single gluon exchange.
In a purely Coulombic potential $V(r) = - (4/3) \alpha_s/r$ the energy levels
are given by $E_n = - [(4/3)\alpha_s]^2 \mu/(2 n^2)$.  In the simplest
approximation the P-wave excitation energy is given by $\Delta E_{P-S} = E_2 -
E_1$.  We have subtracted the S-wave binding energy $B$ from this P-wave
excitation energy to obtain the residual energy difference
\beq
\Delta E_R = \Delta E_{P-S} - B~~.
\eeq
In our convention this S-wave binding energy is positive. In this
simple example it is just the minus the ground-state energy, $-E_1$.
The upshot is that here the residual excitation energy is just the
energy eigenvalue of the P-wave:
\beq
\Delta E_R = E_2 - E_1 - B = E_2 - E_1 + E_1 = E_2 = - [(4/3)\alpha_s]^2\mu/8~.
\eeq
So in this case the slope in Fig.\ \ref{fig:lin} is just
$-[(4/3)\alpha_s]^2 \mu /8$.

In a more realistic potential with a confining piece the slope will be
different and there is likely to be also a constant term.
For light quarks ($u,d,s$) the use of constituent-quark masses means
that it is not necessary to subtract a $B$ term; the constituent-quark
masses already embody such a term.  Nonetheless, the negative slope in
the relation between residual energy and reduced mass is generic. It just
reflects the fact that the P-wave energy (as opposed to energy splitting) is
negative.


This is surprising, as relativistic corrections (important even for
systems as heavy as bottomonium) do not depend purely on the reduced
mass.  This is true for quantum electrodynamics, as shown by the
comparison between positronium and the hydrogen atom \cite{Fulton:1954zz}.
The linear dependence of residual energy must be the result of compensating
effects, not some fundamental relation.  What we have done is to construct a
phenomenological ``bridge'' between confinement and short-distance Coulomb-like
behavior.  This picture then explains why the $B$ and $D$ mesons
are outliers.  Their radii are of order $1/\Lambda_{QCD}$, rather than
$1/(\alpha_s\mu)$.  The fact that $\alpha_s$ runs between $\mu=500$ MeV and
2500 MeV will make the slope slightly scale dependent.

The potential for learning about P-wave excitations of heavy-quark baryons and
mesons makes the present discussion timely.  Consequences have been noted for
charmed and bottom-flavored baryons and mesons.  It will be interesting to
see if some of these regularities are further supported by experiment.

\section*{Acknowledgements}

We thank Sheldon Stone and Tomasz Skwarnicki 
for awakening our interest in P-wave excitations of
heavy-quark systems and for helpful comments.

\end{document}